\documentclass{JHEP}
\usepackage{epsfig}

\newcommand{\be}{\begin{equation}}
\newcommand{\ee}{\end{equation}}

\newcommand{\bea}{\begin{eqnarray}}
\newcommand{\eea}{\end{eqnarray}}
\newcommand{\bean}{\begin{eqnarray*}}
\newcommand{\eean}{\end{eqnarray*}}

\def\beq{\begin{equation}}

\def\eeq{\end{equation}}

\def\Tr{\mathop{\rm Tr}}

\preprint{ hep-th/0212010}

\title{Geometric Dual and Matrix Theory for $SO/Sp$ Gauge Theories}

\author{  
 Bo Feng \\ Institute for Advanced Study \\
Einstein Drive, \\
Princeton, New Jersey, 08540\\
email:~~fengb@ias.edu, 
}

\abstract{In this paper, we give a proof of the equivalence of ${\cal N}=1$
$SO/Sp$ gauge theories deformed from ${\cal N}=2$ by the superpotential
of adjoint field $\Phi$, the dual type IIB superstring theory 
on CY threefold geometries with fluxes and orientifold action
after geometric transition. Furthermore, by relating the geometric
picture to the matrix model, we show the equivalence between the field theory
and the corresponding matrix model. }

\begin{document}

\section{Introduction}
A few months ago, a deep relationship between matrix models 
and supersymmetric gauge field theories has been pointed out in 
\cite{Dijkgraaf:2002fc,Dijkgraaf:2002vw,Dijkgraaf:2002dh}. In these
papers, it was shown that  exact glueball effective actions
for supersymmetric gauge field theories can be calculated by 
planar diagrams of  corresponding matrix models. Since then,
a lot of works have been done to  check this conjecture
by explicit examples (e.g. \cite{Dorey:2002jc} to \cite{Suzuki:2002gp},
and  \cite{Ferrari:2002kq} to \cite{Tachikawa:2002ud}),
from which some remarkable features of the new method have been 
demonstrated. For example,  in \cite{Dorey:2002jc,Dorey:2002tj,Dorey:2002pq,
Dijkgraaf:2002pp} it was shown that different massive vacua of
the mass deformed ${\cal N}=4$ theory are related to each other by
 $SL(2,Z)$ modular groups, so the Montonen-Olive duality is not
an assumption, but rather a derived result. It was also shown that
the matrix model can calculate not only the exact low energy 
superpotential, but also  quantum corrections of  classical
moduli spaces.

Among these results, two papers \cite{Dijkgraaf:2002xd} and
\cite{Cachazo:2002ry} used purely  the language of field theories
to prove the DV conjecture. These two proofs are very useful
because they do not rely on the geometric picture and explain why
 calculations of effective actions can be reduced to matrix
models. They provide also bases to generalize to other
interesting cases, for example, the double trace deformation
studied in \cite{double} or the $SO/Sp$ gauge groups studied in
\cite{Ita:2002kx}.

With these developments, it seems that to prove the DV conjecture,
the geometric picture is not really needed. However, the field
theory proof is not very general at this moment and for theories
which can be geometrically engineered and embedded into string theory,
the geometric method has been proven to be a very useful alternative. 
One  explicit 
example can be found in \cite{Dijkgraaf:2002fc} for the ${\cal N}=1$ 
$U(N)$ theory with one adjoint field $\Phi$ and arbitrary 
superpotential $W=\sum_{r=1}^{n+1} g_r u_r=
\sum_{i=1}^{n+1} g_r \Tr(\Phi^r)/r$. The proof in \cite{Dijkgraaf:2002fc}
was based on works
of \cite{Cachazo:2001jy,Cachazo:2001gh,Cachazo:2001sg,Cachazo:2002pr}
where it was shown by large N duality that the calculation in 
the field theory
is equivalent to the one in the dual geometry. So if we can derive
the dual geometry (the spectral curve and periods of cycles) 
 from the matrix model, by the link between the geometry and field
theory, the relationship between the matrix model and field theory
is established also.

In this paper, we will use the same logic
to extend the proof of DV conjecture to ${\cal N}=1$ 
$SO(N)/Sp(2N)$ theories with one adjoint field $\Phi$ and arbitrary 
superpotential $W=\sum_{r=1}^{n+1} g_{2r} u_{2r}=
\sum_{r=1}^{n+1} g_{2r} \Tr(\Phi^{2r})/2r$. We will show first that
the exact effective action calculated by field theory method is same
as the one calculated by the dual  geometry method. 
Then we derive the corresponding spectral curve
from the  matrix model and match physical quantities such as  $S_i$
and $\Pi_i$ at two sides of the matrix model and dual geometry.
Combining the first step, it will complete our proof of DV conjecture
for $SO/Sp$ gauge groups.

The organization of the paper is following. In section two we 
provide the analysis in field theory.
In section three we review the geometric
dual picture and give a  proof of the equivalence between the
gauge theory and dual geometry. 
In section four, we present the derivation of the dual geometry 
from  the matrix model, thus
 close the loop of our proof.\footnote{When  submitting
this paper, we noticed that two papers \cite{Yutaka} and \cite{Sujay}
have some overlaps with this paper.}

\section{The  analysis in field theory}
First let us analyze the classical moduli space of 
$SO(2N)$, $SO(2N+1)$ and $Sp(2N)$ (the notation for $Sp(2N)$ is that
the rank of the gauge group is $N$) with  following superpotential 
\be
\label{def_sup} W=\sum_{r=1}^{n+1} g_{2r} u_{2r}=
\sum_{r=1}^{n+1} g_{2r} {\Tr(\Phi^{2r})\over 2r}~.
\ee
By gauge transformations, we can rotate the  $\Phi$ into following form:
$diag(x_1 i\sigma_2,..., x_N i\sigma_2)$ for $SO(2N)$,
$diag(x_1 i\sigma_2,..., x_N i\sigma_2,0)$ for $SO(2N+1)$ and
$diag(x_1,-x_1,...,x_N,-x_N)$ for $Sp(2N)$ with $\sigma_2$ the Pauli
matrix.
The supersymmetric vacua are given by  solutions of  F-terms, i.e., 
roots of 
\be
\label{F-term}
W'(x)= g_{2n+2} x \prod_{j=1}^n (x^2 \pm a_i^2),~~~\pm~~for~~SO/Sp.
\ee 
If we choose $N_i$ $x_i$ to be the same value $ia_i(a_i)$ (with the 
convention that $a_0=0$), the gauge group is broken to
\be
\label{broken} 
SO(2N) \rightarrow SO(2N_0)
\prod_{j=1}^n U(N_j)
\ee
with $\sum_{j=0}^n N_j=N$ for $SO(2N)$ (for $SO(2N+1)$ or $Sp(2N)$,
$SO(2N_0)$ is replaced by $SO(2N_0+1)$ or $Sp(2N_0)$). At low energy, 
$SO(2N_0)$, $SO(2N_0+1)$, $Sp(2N_0)$
as well as $SU(N_i)$ develop a mass gap and confine, so there are
$n$ massless $U(1)$ gauge fields left. It is the exact effective
action for these fields we are looking for.

Since our theories are  deformed from  corresponding 
 ${\cal N}=2$ theories  by the superpotential (\ref{def_sup}),
we can calculate the exact superpotentials for these deformed theories
by using the well-known Seiberg-Witten curves \cite{Seiberg:1994rs,
Seiberg:1994aj,Klemm:1994qs,Argyres:1994xh,Hanany:1995na,Argyres:1995wt,
Brandhuber:1995zp,Danielsson:1995is,Hanany:1995fu,Argyres:1995fw,
D'Hoker:1996mu}. The method has been elaborated in \cite{Cachazo:2001jy,
Edelstein:2001mw}.

The basic idea is that   ${\cal N}=2$ theories deformed only by
$W$ of (\ref{def_sup})  have unbroken supersymmetries on a submanifold
of  Coulomb branches, where there are additional $l$ massless fields
besides $u_r$, such as magnetic monopoles or dyons. The exact
low energy superpotential in these vacua is given  by
\be
\label{def_sup_exact} W_{eff}=\sum_{r=1}^{n+1} g_{2r} \langle u_{2r}
\rangle
\ee 
with $\langle u_{2r}\rangle$ taking value  in the submanifold
where $l$ monopoles are massless. In other words, we require
$l$ mutually local monopoles or dyons in the  submanifold
of  Coulomb branches. This requirement put $l$ conditions in  original
Coulomb branches and $\langle u_{2r}\rangle$ lie on the
codimension $l$ submanifold. 

Because  $\langle u_{2r}\rangle$ lie on the
codimension $l$ submanifold, we can parameterize them by $(N-l)$ parameters.
To get the low energy effective action, we need to minimize $W_{eff}$ in
(\ref{def_sup_exact}) regarding  these parameters and substitute
results 
back to $W_{eff}$. By this way, we get the low energy effective action 
$W_{low}$ as
a function of $g_{2r}$ and $\Lambda$ only.

The above conditions can be translated into the requirement of
 proper factorization forms of corresponding Seiberg-Witten curves
as shown in \cite{Cachazo:2001jy,Edelstein:2001mw}. For 
 $SO/Sp$ gauge groups, as remarked in \cite{Edelstein:2001mw},
there are two forms of  SW curves. One is as a hyperelliptic
curve of genus $N$ in \cite{Argyres:1995fw} and another is as 
a hyperelliptic  curve of genus $2N$ with $Z_2$ symmetry in
\cite{Danielsson:1995is,Brandhuber:1995zp,D'Hoker:1996mu}.
It was found that to connect to the geometric picture, the second
choice is  more natural and will be used throughout  the paper.

Let us see how it works by the example of $SO(2N)$. $SO(2N)$ can be embedded
into $U(2N)$ and considered as the $Z_2$ quotient of later. With
the superpotential (\ref{def_sup}) (\ref{F-term}), $U(2N)$ is broken to
$2n+1$ factors as
$$
U(2N)\rightarrow U(2N_0) \prod_{j=1}^n U(N_{j+}) \times U(N_{j-})
$$
with $2N_0+\sum_{j=1}^n (N_{j+}+N_{j-})=2N$ and the 
corresponding SW curve is factorized
as
$$
y^2=F_{2(2n+1)}(x) H_{2N-(2n+1)}(x)^2~.
$$
However, to reduce to $SO(2N)$ group, we must take the $Z_2$ action which
requires $N_{j+}=N_{j-}$. The $Z_2$ action also maps $ U(N_{j+})$ 
located at $ia_j$ to 
$ U(N_{j-})$ located at $-ia_j$ 
 and projects $U(2N_0)$ located at $0$ to $SO(2N_0)$, so 
finally we get the
breaking pattern
$$
SO(2N)\rightarrow SO(2N_0)\prod_{j=1}^n U(N_j).
$$
Considering the $Z_2$ action of the factorized SW curve, we get
\cite{Edelstein:2001mw}
$$
y^2  =  P_{2N}(x^2,u)^2- 4\Lambda^{4N-4} x^4 = (xH_{2N-(2n+2)}(x))^2
  F_{2(2n+1)}(x),
$$
where  both $H(x)$ and $F(x)$ are functions of $x^2$.
Knowing the factorized form,  the gauge coupling constants of
the remaining massless $U(1)$ fields can be calculated by the period matrix
of the reduced curve
\be
y^2= F_{2(2n+1)}(x^2; \langle u_{2r} \rangle)= F_{2(2n+1)}(x^2; g_{2r},
\Lambda)
\ee

As we will show shortly, the function $F_{2(2n+1)}(x^2)$ is related to
the deformed superpotential and geometry by
$$
g_{2n+2}^2 F_{2(2n+1)}(x^2)= W'(x)^2+ f_{2n}(x)
$$
where $f_{2n}(x)$ with degree $2n$ is a function of $x^2$.

\subsection{Rephrasing the   problem}
As shown in \cite{Cachazo:2001jy,Cachazo:2002pr}, the factorization 
and extremum
can be restated into a pure algebraic 
problem which is well posed and has a unique solution: {\sl 
Find $P_{2N}(x; u)$ such that\footnote{Following discussions
are under the assumption that the wrapping number $N_i\neq 0$
for all $i=0,...,n$ which is also used in \cite{Cachazo:2002pr}.
The discussion for more general cases is  under  investigation.} 
\bea
SO(2N): & ~~~ & P_{2N}(x^2,u)^2-4\Lambda^{4N-4} x^4= 
x^2 H^2_{2N-2n-2}(x)F_{2(2n+1)}(x)  \nonumber \\
& & =x^2 H^2_{2N-2n-2}(x) {1\over g^2_{2n+2}} (W'(x)^2+f_{2n}(x)), \\
SO(2N+1): & ~~~ & P_{2N}(x^2,u)^2-4\Lambda^{4N-2} x^2=x^2 H^2_{2N-2n-2}(x)
F_{2(2n+1)}(x)  \nonumber \\
& & = x^2 H^2_{2N-2n-2}(x)
{1\over g^2_{2n+2}} (W'(x)^2+f_{2n}(x)), \\
Sp(2N): & ~~~ & [x^2P_{2N}(x^2,u)+2\Lambda^{2N+2}]^2-4\Lambda^{4N+4} 
=x^2 H^2_{2N-2n}(x) F_{2(2n+1)}(x)  \nonumber \\
& & = x^2 H^2_{2N-2n}(x){1\over g^2_{2n+2}} (W'(x)^2+f_{2n}(x)), 
\eea
where $W'(x)= g_{2n+2}x\prod_{i=1}^n (x^2\pm a_i^2)$ (where $+$ for $SO$ and
$-$ for $Sp$) is given, together
with following boundary conditions at $\Lambda \rightarrow 0$ as
\bea
SO(2N): & ~~~ & P_{2N}(x^2,u) \rightarrow x^{2N_0} \prod_{i=1}^n
  (x^2+a_i^2)^{N_i},~~~~~\sum_{i=0}^n N_i=N, \\
SO(2N+1): & ~~~ & P_{2N}(x^2,u) \rightarrow x^{2N_0} \prod_{i=1}^n
  (x^2+a_i^2)^{N_i},~~~~~\sum_{i=0}^n N_i=N, \\
Sp(2N): & ~~~ & P_{2N}(x^2,u) \rightarrow x^{2N_0} \prod_{i=1}^n
  (x^2-a_i^2)^{N_i},~~~~~\sum_{i=0}^n N_i=N, 
\eea
}
Above boundary conditions mean that  gauge groups are broken as
\bean
SO(2N) & \rightarrow  & SO(2N_0) \times \prod_{i=1}^n U(N_i), \\
SO(2N+1) & \rightarrow  & SO(2N_0+1) \times \prod_{i=1}^n U(N_i), \\
Sp(2N) & \rightarrow  & Sp(2N_0) \times \prod_{i=1}^n U(N_i). 
\eean
Using  same method as in 
\cite{Cachazo:2002pr} it can be proved that  solutions for
above problems are unique.

Once the low energy effective action 
$$ W_{low}(g_{2r},\Lambda)= \sum_{r=1}^{n+1} g_{2r} \langle u_{2r} 
\rangle $$
is obtained, we can calculate 
\bea
{\partial W_{eff} \over \partial g_{2r}} & = & \langle u_{2r} \rangle,\\
{\partial W \over \partial \log(\Lambda^{2\hat{N}})} & = & {-b_{2n} 
\over 4g_{2n+2}} \label{leading}.
\eea
where $\hat{N}$ is $2N-2$ for $SO(2N)$, $2N-1$ for $SO(2N+1)$ and $2N+2$ for
$Sp(2N)$. The $S_0$ is the glueball superfield for $SO(2N_0)$, $SO(2N_0+1)$
or $Sp(2N_0)$ factor and $S_i$ is the glueball superfield for $U(N_i)$
factor.
The $b_{2n}$ is the leading coefficient of the function $f_{2n}(x)$. We will
give  derivations of these results in next subsection.

\subsection{The function $F_{2(2n+1)}(x)$}
As we mentioned above, the function $F_{2(2n+1)}(x)$ is related to
the deformed superpotential and geometry by 
\be \label{fun_F}
g_{2n+2}^2 F_{2(2n+1)}(x^2)= W'(x)^2+ f_{2n}(x)~.
\ee
This result has been given in \cite{Edelstein:2001mw} for $SO$ groups.
Here we adopt the method in \cite{Cachazo:2001jy,Cachazo:2002pr} which will also enable us
to show relation (\ref{leading}).

Let us start with the $SO(2N)$ gauge group. In this case, 
the SW curve is factorized as
$$
 P_{2N}(x^2,u)^2- 4\Lambda^{4N-4} x^4 = (\widetilde{H}_{2N-2n-1}(x))^2
 F_{2(2n+1)}(x)=
(xH_{2N-(2n+2)}(x))^2
  F_{2(2n+1)}(x)
$$
Notice  that since both the left hand side and $ F_{2(2n+1)}(x)$ are functions
of $x^2$ and the degree of $\widetilde{H}_{2N-2n-1}(x)$ is odd, one
factor $x$ must be factorized out in $\widetilde{H}_{2N-2n-1}(x)$, 
thus we can write $\widetilde{H}_{2N-2n-1}(x)=xH_{2N-(2n+2)}(x)$. 
For our convenience,  we change it
to 
 \be
({P_{2N}(x^2,u) \over x^2})^2- 4\Lambda^{4N-4}  = (H_{l}(x))^2 x^{-2}
  F_{4N-2l-2}(x),
\ee
with ${P_{2N}(x^2,u) \over x^2}$  a polynomial of $x^2$. 
As in \cite{Cachazo:2002pr}, the problem of factorizing the SW curve and 
minimizing  the superpotential under these constraints can be translated into
minimizing the following superpotential 
\be
\label{WmultiSO2N} W= \sum_{r=1}^{n+1} g_{2r} u_{2r}+\sum_{i=1}^l 
[L_i (({P_{2N}(x^2,u) \over x^2})|_{x=p_i}-2\epsilon_i \Lambda^{2N-2} )
+ Q_i 
{\partial   ({P_{2N}(x^2,u) \over x^2}) \over \partial x}|_{x=p_i}]
\ee  
with $\epsilon_i=\pm 1$ and variables $u_{2r}$ and Lagrange multipliers
$L_i,Q_i,p_i$. In fact $L_i,Q_i$ conditions tell us that there are $l$ double
roots at $p_i$ as shown by the factor  $(H_{l}(x))^2$. 

From the equation (\ref{WmultiSO2N}) we first get
\bean
{\partial \over \partial Q_i}: & ~~~ & 
{\partial   ({P_{2N}(x^2,u) \over x^2})  \over \partial x}|_{x=p_i}=0 \\
{\partial \over \partial p_i}: & ~~~ &  Q_i 
{\partial^2   ({P_{2N}(x^2,u) \over x^2})   \over \partial x^2}|_{x=p_i}
=0. \\
\eean
Since in general ${\partial^2   ({P_{2N}(x^2,u) \over x^2})
\over \partial x^2}$ is not degenerate, we get $Q_i=0$. Using 
this result we get
\bean
{\partial \over \partial u_{2r}}:  & ~~~ & 
 g_{2r}+\sum_{i=1}^l \sum_{j=0}^{N} L_i p_i^{2N-2j-2} {\partial s_{2j} \over
 \partial u_{2r}}=0, \\
& \rightarrow & g_{2r}=\sum_{i=1}^l \sum_{j=0}^{N} L_i p_i^{2N-2j-2}
s_{2j-2r}
\eean
where we have used the expansion $P_{2N}(x^2,u)=\sum_{r=0}^{N}
s_{2r} x^{2N-2r}$ with $s_0=1$ and  
${\partial s_{2j} \over \partial u_{2r}}=-s_{2j-2r}$.
Because the SW curve is an even function of $x$, roots 
$p_i$ must be in pairs as $(p_i,-p_i)$ and we can write the sum as
\be
g_{2r}=\sum_{i=1}^{l/2} \sum_{j=0}^{N} (L_{i+}+L_{i-}) p_i^{2N-2j-2}
s_{2j-2r}
\ee
where we have assumed that $l$ is even number.

Now we calculate 
\bean
W'(x) & = & \sum_{r=1}^{N} g_{2r} x^{2r-1}  \\
 & = & \sum_{r=1}^{N} \sum_{i=1}^{l/2} \sum_{j=0}^{N} 
(L_{i+}+L_{i-}) p_i^{2N-2j-2}
s_{2j-2r} x^{2r-1}  \\
 & = & \sum_{r=-\infty}^{N} \sum_{i=1}^{l/2} \sum_{j=0}^{N}
(L_{i+}+L_{i-}) p_i^{2N-2j-2}
s_{2j-2r} x^{2r-1} -2L \Lambda^{2N-2} x^{-1}+  {\cal O}(x^{-3}) \\
\eean
where the $L\equiv \sum_{i=1}^{l/2} (L_{i+}+L_{i-})\epsilon_i$.
It can be shown by taking  ${\partial \over \partial L_i}$ of
(\ref{WmultiSO2N}) 
that $\epsilon_{i}=\epsilon_{i+}=\epsilon_{i-}$. Replacing
$\sum_{j=0}^{N}$ by $\sum_{j=-\infty}^{N}$ since these  terms are
of  higher order, we get
\bean
W'(x) & = & \sum_{r=-\infty}^{N} \sum_{i=1}^{l/2} \sum_{j=-\infty}^{N}
(L_{i+}+L_{i-}) p_i^{2N-2j-2}
s_{2j-2r} x^{2r-1} -2L \Lambda^{2N-2} x^{-1}+  {\cal O}(x^{-3}) \\
& = & \sum_{i=1}^{l/2} \sum_{j=-\infty}^{N}(L_{i+}+L_{i-}) p_i^{2N-2j-2}
x^{-2N-1+2j} \sum_{\widetilde{r}\equiv j-r: j-N}^{+\infty} s_{2\widetilde{r}}
x^{2N-2\widetilde{r}} -2L \Lambda^{2N-2} x^{-1}+  {\cal O}(x^{-3}) \\
& = &P_{2N}(x^2,u)  \sum_{i=1}^{l/2} \sum_{j=-\infty}^{N}(L_{i+}+L_{i-}) p_i^{2N-2j-2}
x^{-2N-1+2j}  -2L \Lambda^{2N-2} x^{-1}+  {\cal O}(x^{-3}) \\ 
& = &P_{2N}(x^2,u)  \sum_{i=1}^{l/2} {(L_{i+}+L_{i-}) \over x p_i^2}
{1\over 1-{p_i^2 \over x^2}}  -2L \Lambda^{2N-2} x^{-1}+  {\cal O}(x^{-3}) \\ 
& = & xP_{2N}(x^2,u)  \sum_{i=1}^{l/2} {(L_{i+}+L_{i-}) \over  p_i^2}
{1\over x^2- p_i^2}  -2L \Lambda^{2N-2} x^{-1}+  {\cal O}(x^{-3})~. \\ 
\eean
With the definition 
$$
 \sum_{i=1}^{l/2} {(L_{i+}+L_{i-}) \over  p_i^2 } 
{x^2 \over x^2- p_i^2} = {B_{l} (x) \over H_{l}(x)}
$$
we have
\be \label{wprime}
W'(x)  =    B_{l}(x) {P_{2N}(x^2,u) \over x H_{l}(x)} -2L \Lambda^{2N-2} x^{-1}+  {\cal O}(x^{-3})~. \\ 
\ee
From this we can write
\bean
W'(x) +2L \Lambda^{2N-2} x^{-1}=  B_{l}(x) \sqrt{  F_{4N-2l-2}(x)+
{4\Lambda^{4N-4} x^2 \over H_l(x)^2}}+{\cal O}(x^{-3})~. \\
\eean
Comparing the degree at two sides we find $deg(B_l)=2n+1-(2N-l-1)\geq 0$.
If we set $l=2N-2n-2$, $B_l(x)=g_{2n+2}$ is just a constant, so finally
we get the wanted relationship
\be
\label{coeffSO2N}
g_{2n+2}^2 F_{2(2n+1)}(x)= W^{'2}(x)+ 4 L \Lambda^{2N-2} g_{2n+2} x^{2n}+
...=W^{'2}(x)+f_{2n}(x)
\ee 
Furthermore from the form (\ref{wprime}), it can be seen that both 
$H_l(x)$ and $F_{4N-2l-2}(x)$ are functions of $x^2$.

There is another important relationship we can get. Differentiating 
(\ref{WmultiSO2N}) regarding to  $\log(\Lambda^{4N-4})$, we have
\bean
{d W \over d \log(\Lambda^{4N-4})} 
& = & {\partial W \over \partial \log(\Lambda^{4N-4})} +
{\partial W \over \partial u_{2r}}{\partial u_{2r} \over \partial
\log(\Lambda^{4N-4})}\\ & & +{\partial W \over \partial L_i}{\partial L_i
\over \partial\log(\Lambda^{4N-4})}+ 
{\partial W \over \partial p_i}{\partial p_i
\over \partial\log(\Lambda^{4N-4})} \\
& = & {\partial W \over \partial \log(\Lambda^{4N-4})} 
=  - \sum_{i}^{l} L_i\epsilon_i \Lambda^{2N-2} \\
&  = &   - L \Lambda^{2N-2}
\eean
where in the third line we have used  equations for $u_{2r},p_i,L_i$
and in the fourth line, the definition of $L$.  
From (\ref{coeffSO2N}) we can read out  the leading coefficient of
$f_{2n}(x)$ to be $b_{2n}=4 L \Lambda^{2N-2} g_{2n+2}$, so we get
\be
\label{W_scale_SOe}
{d W \over d \log(\Lambda^{4N-4})} =-{b_{2n} \over 4
g_{2n+2}} 
\ee
which has been advertised in (\ref{leading}).

\subsubsection{The $SO(2N+1)$ and $Sp(2N)$ cases}
Having done the case of $SO(2N)$ in detail, we will just scratch
the $SO(2N+1)$ and $Sp(2N)$ cases. For $SO(2N+1)$, we write the
factorized SW curve  as 
\be
({P_{2N}(x^2,u) \over x})^2- 4\Lambda^{4N-2}  = (H_{l}(x))^2 
  F_{4N-2l-2}(x),
\ee
and the corresponding low energy effective action
\be
\label{WmultiSO2N1} W= \sum_{r=1}^{n+1} g_{2r} u_{2r}+\sum_{i=1}^l 
[L_i (({P_{2N}(x^2,u) \over x})|_{x=p_i}-2\epsilon_i \Lambda^{2N-1} )
+ Q_i 
{\partial   ({P_{2N}(x^2,u) \over x}) \over \partial x}|_{x=p_i}]
\ee
Using  equations of $Q_i,L_i,p_i,u_{2r}$ we get
\bean
g_{2r} & = & \sum_{i=1}^{l/2} \sum_{j=0}^{N} (L_{i+}-L_{i-}) p_i^{2N-2j-1}
s_{2j-2r}
\eean
and
\bean
W'(x) & = & \sum_{r=1}^{N} g_{2r} x^{2r-1}  \\
& = & \sum_{r=-\infty}^{N} \sum_{i=1}^{l/2} \sum_{j=0}^{N}
(L_{i+}-L_{i-}) p_i^{2N-2j-1}
s_{2j-2r} x^{2r-1} -2L \Lambda^{2N-1} x^{-1}+  {\cal O}(x^{-3}) \\
& = & xP_{2N}(x^2,u)  \sum_{i=1}^{l/2} {(L_{i+}-L_{i-}) \over  p_i}
{1\over x^2- p_i^2}  -2L \Lambda^{2N-1} x^{-1}+  {\cal O}(x^{-3}) \\ 
\eean
with the definition $L\equiv \sum_{i=1}^{l/2} (L_{i+}-L_{i-})\epsilon_i$.
Defining  
$$\sum_{i=1}^{l/2} {(L_{i+}-L_{i-}) \over  p_i } 
{x^2 \over x^2- p_i^2} = {B_{l} (x) \over H_{l}(x)}$$
we can write 
\bean
W'(x) +2L \Lambda^{2N-1} x^{-1}=  B_{l}(x) \sqrt{  F_{4N-2l-2}(x)+
{4\Lambda^{4N-2}  \over H_l(x)^2}}+{\cal O}(x^{-3}). \\
\eean
Setting $l=2N-2n-2$, $B=g_{2n+2}$ we finally get 
\be
\label{coeffSO2N1}
g_{2n+2}^2 F_{2(2n+1)}(x)= W^{'2}(x)+ 4 L \Lambda^{2N-1} g_{2n+2} x^{2n}+
...=W^{'2}(x)+f_{2n}(x)
\ee 
Again, differentiating (\ref{WmultiSO2N1}) by $\log(\Lambda^{4N-2})$,
we get
\be  \label{W_scale_SOo} 
{d W \over d \log(\Lambda^{4N-2})} 
 = - \sum_{i}^{l} L_i\epsilon_i \Lambda^{2N-1}  =  - L \Lambda^{2N-1}
=-{b_{2n} \over 4 g_{2n+2}} 
\ee

For the $Sp(2N)$ gauge group  we write down
\be
[x^2 P_{2N}(x^2,u)+ 2\Lambda^{2N+2}]^2-4\Lambda^{4N+4}
  = (H_{l}(x))^2
  F_{4N-2l+2}(x) x^2, 
\ee
and 
\be
\label{WmultiSp2N} W= \sum_{r=1}^{n+1} g_{2r} u_{2r}+\sum_{i=1}^l 
[L_i (x^2 P_{2N}(x^2,u)|_{x=p_i}-2\epsilon_i \Lambda^{2N+2} )
+ Q_i 
{\partial  (x^2 P_{2N}(x^2,u))  \over \partial x}|_{x=p_i}]
\ee
where $\epsilon_i=0,-2$ which is different from the $SO$ case. 
From the $W$, we find 
\bean
g_{2r}= \sum_{i=1}^{l} \sum_{j=0}^N L_i p_i^{2N-2j+2} s_{2i-2r}
=\sum_{i=1}^{l/2} \sum_{j=0}^N (L_{i+}+L_{i-}) p_i^{2N-2j+2} s_{2i-2r}
\eean
and 
\bean
W'(x) & = & \sum_{r=1}^{N} g_{2r} x^{2r-1}  \\
& = & \sum_{r=-\infty}^{N} \sum_{i=1}^{l/2} \sum_{j=0}^{N}
(L_{i+}+L_{i-}) p_i^{2N-2j+2}
s_{2j-2r} x^{2r-1} -2L \Lambda^{2N+2} x^{-1}+  {\cal O}(x^{-3}) \\
& = & xP_{2N}(x^2,u)  \sum_{i=1}^{l/2} {(L_{i+}+L_{i-})p_i^2 }
{1\over x^2- p_i^2}  -2L \Lambda^{2N+2} x^{-1}+  {\cal O}(x^{-3}) \\ 
\eean
with the definition $L\equiv \sum_{i=1}^{l/2} (L_{i+}+L_{i-})\epsilon_i$.
Writing 
$$
\sum_{i=1}^{l/2} {(L_{i+}+L_{i-}) p_i^2}{1\over x^2- p_i^2}={B_{l-2}(x) \over
H_l(x)}
$$
we get
 \bean
W'(x) & = & B_{l-2}(x) {1\over x}{x^2 P_{2N}(x^2,u) \over H_l(x)} -2L \Lambda^{2N+2} x^{-1}+  {\cal O}(x^{-3}) \\
  & = & B_{l-2}(x) {1 \over x} 
(\sqrt{ F_{4N-2l+2}(x) x^2+{4 \Lambda^{4N+4} \over H_l(x)^2}}
-{2 \Lambda^{2N+2} \over H_l(x)})-2L \Lambda^{2N+2} x^{-1}
+ {\cal O}(x^{-2}) \\ 
& = & B_{l-2}(x) (\sqrt{ F_{4N-2l+2}(x) +{4 \Lambda^{4N+4} \over x^2 H_l(x)^2}}
-{2 \Lambda^{2N+2} \over x H_l(x)})-2L \Lambda^{2N+2} x^{-1}
+ {\cal O}(x^{-2}) \\ 
\eean
Setting $l=2N-2n$ and $B_{l-2}(x)=g_{2n+2}$, we have
\be
\label{coeffSp2N}
g_{2n+2}^2 F_{2(2n+1)}= W'(x)^2+ 4 L \Lambda^{2N+2} g_{2n+2} x^{2n}+...=
W'(x)^2+f_{2n}(x)
\ee
Differentiating (\ref{WmultiSp2N}) by $\log(\Lambda^{4N+4})$ we found 
\be  \label{W_scale_Sp}
{d W \over d \log(\Lambda^{4N+4})} 
 = - \sum_{i}^{l} L_i\epsilon_i \Lambda^{2N+2}  =  - L \Lambda^{2N+2}
=-{b_{2n} \over 4 g_{2n+2}} 
\ee

\subsection{The $\Lambda \rightarrow 0$ limit}
To compare with the geometric picture, we need to discuss the solution in 
field theory at the limit $\Lambda \rightarrow 0$. In this
limit, the factorization is easy to be solved. For example, for 
$SO(2N)$ gauge groups, we propose that  
\bean
 [P_{2N}(x,u)]^2 & = &  [x^{2N_0} \prod_{i=1}^n (x^2+ a_i^2)^{N_i}]^2 \\
& = & [{W'(x)^2 \over g_{2n+2}^2}] x^2[x^{2N_0-2} 
\prod_{i=1}^n (x^2+ a_i^2)^{N_i-1}]^2 
\eean
where the first line tells us that 
the proposed $P_{2N}(x,u)$ does satisfy the boundary condition.
From this factorized form we can read out that $f_{2n}(x)=0$.  
In this limit  the
effective action is calculated  as
\bea
W_{eff} & = & \sum_{r=1}^{n+1} {g_{2r} \over 2r} \Tr[\Phi^{2r}] 
 =  \sum_{r=1}^{n+1} {g_{2r}\over 2r}
 \sum_{i=-n}^{n} (-)^r N_i a_i^{2r}  \nonumber\\
& = &  \sum_{i=-n}^{n} N_i  \sum_{r=1}^{n+1} {g_{2r}\over 2r} (-)^r
a_i^{2r} \label{W_field_SO}= \sum_{i=-n}^{n} N_i W(\alpha_i)
\eea
where we have used the from 
 $\Phi=diag( 0_{2N_0}, (i a_i,-i a_i)_{N_i})$ and $\alpha_i$ are these
eigenvalues.

Similar calculations can be done for other two gauge groups as
\bean
SO(2N+1): & ~~~ & [P_{2N}(x,u)]^2  = 
 [x^{2N_0} \prod_{i=1}^n (x^2+ a_i^2)^{N_i}]^2 \\
&  &= [{W'(x)^2 \over g_{2n+2}^2}] x^2[x^{2N_0-2} 
\prod_{i=1}^n (x^2+ a_i^2)^{N_i-1}]^2, \\
Sp(2N): & ~~~ & x^4  [P_{2N}(x,u)]^2  = x^4 
[x^{2N_0} \prod_{i=1}^n (x^2- a_i^2)^{N_i}]^2 \\
& & = [{W'(x)^2 \over g_{2n+2}^2}] x^2 [x^{2N_0} 
\prod_{i=1}^n (x^2+ a_i^2)^{N_i-1}]^2, \\
\eean
with $f_{2n}(x)=0$. The effective action of $SO(2N+1)$ is same
as $SO(2N)$, but for $Sp(2N)$ it is modified to
\bean
W_{eff} & = & \sum_{r=1}^{n+1} {g_{2r} \over 2r} Tr[\Phi^{2r}] 
 =  \sum_{r=1}^{n+1} {g_{2r}\over 2r}
 \sum_{i=-n}^{n}  N_i a_i^{2r} \\
& = &  \sum_{i=-n}^{n} N_i  \sum_{r=1}^{n+1} {g_{2r}\over 2r} 
a_i^{2r}=  \sum_{i=-n}^{n} N_i W(\alpha_i)\\
\eean
where  
 $\Phi=diag( 0_{2N_0}, (a_i, -a_i)_{N_i})$.

\section{The geometric picture}
The  geometric duals to field theories are given in \cite{Cachazo:2001jy},
where it was conjectured that low energy (holomorphic) dynamics can be 
calculated by geometric dual theories. Later in \cite{Cachazo:2002pr},
this conjecture has been proved for  $U(N)$ gauge groups with one adjoint
field $\Phi$. The geometric dual theories have been generalized from 
$U(N)$ gauge groups to $SO/Sp$ gauge groups 
in \cite{Edelstein:2001mw} and explicit examples
to support this conjecture were given in \cite{Fuji:2002vv}.
It is our aim in this paper to give a proof for  $SO/Sp$ gauge groups.

\subsection{Review}
To have the geometric dual theory, first we need to geometrically engineer
the ${\cal N}=2$ field theory deformed by  superpotential (\ref{def_sup}).
It can be done by wrapping D5-branes along   two cycles
in the non-compact, nontrivial fibrated Calabi-Yau threefold
\be \label{sing_geo}
uv+w^2+W'(x)^2=0,~~~~~~W'(x)= g_{2n+2} x \prod_{j=1}^n (x^2 \pm a_i^2),
\ee  
At each root of $W'(x)$  there is a blown up $S^2$ with $N_i$ D5-branes
wrapped around this $S^2$. The geometric dual theory is obtained via 
the geometric transition \cite{Vafa:2000wi,Gopakumar:1998ki} 
where $S^2$'s are blown
down and $S^3$'s are blown up. At the same time, $N_i$ D5-branes wrapped 
around $S_i^2$ disappear and are replaced by $N_i$ units of $H_{RR}$ fluxes
through the new nontrivial $S_i^3$. The transition to $S^3$'s corresponds
to a complex deformation of the geometry as
\be \label{def_geo}
uv+w^2+W'(x)^2+f_{2n}(x)=0,
\ee 
From this, we can calculate the effective superpotential in the geometric
dual theory by
\be \label{geo_eff}
{-1\over 2\pi i} W_{eff} = \int_{CY} H\wedge \Omega= 
\sum_{i=-n}^{n} \int_{A_i} H \int_{B_i} \Omega -\int_{B_i} H
\int_{A_i} \Omega 
\ee
where $H=H_{RR}-\tau_{IIB} H_{NS}$, $\Omega$ the holomorphic three
form on the CY 3-fold and $A_i,B_i$ the symplectic bases.

As did in \cite{Cachazo:2002pr} we can reduce the integration on the
CY 3-fold to the integration on the reduced 
 surface
\be \label{red_Rie}
y^2= W'(x)^2+f_{2n}(x)
\ee
with reduced one forms $h$ and $dx\lambda_{eff}$ 
\bea
dx\lambda_{eff} & = & dx  \sqrt{W'(x)^2+f_{2n}(x)}, \\
h & = & \int_{S^2} H,
\eea
so the effective action is simplified to 
\bea
{-1\over 2\pi i} W_{eff}& = &\int_{\Gamma} h\wedge dx \lambda_{eff} =
\sum_{i=-n}^{n} \int_{a_i} h \int_{b_i} dx\lambda_{eff}
 -\int_{b_i} h
\int_{a_i} dx\lambda_{eff}  \label{reduced_W}
\eea
with $a_i$  these compact one cycles and $b_i$, these corresponding 
non-compact dual one cycles. 

When we discuss the $SO/Sp$ gauge groups, we need to add the orientifold into
the geometry\cite{Sinha:2000ap,Edelstein:2001mw,Landsteiner:1997vd}. 
The orientifold action will have following contributions. Firstly it
contributes  $H_{RR}$ fluxes to the integration along the cycle around
it. Secondly it pairs  blown up $S^3$'s except the one fixed by
the orientifold action in CY 3-fold. 
In other words, the orientifold action requires the
deformation $f_{2n}(x)$ to be a function of $x^2$.

Above discussions provide us with a convenient way to look at the problem.
We can work first at the double covering space, where the result of 
$U(2N)$ can be applied, with the condition that it preserves the
$Z_2$ action of orientifold.  Then by putting the $Z_2$ action back, 
we get the results for $SO/Sp$ gauge groups.

First let us discuss the choice of cycles in (\ref{reduced_W}). 
These cycles are  almost the same as these  
in \cite{Cachazo:2002pr} with  only one extra condition that they
preserve the $Z_2$ symmetry (see Figure  \ref{choice}).
Since  branch cuts are symmetric with one located at the center, 
it is not difficult to show
\be
\label{inte_rel}
\oint_{\alpha_{k}} dx\lambda_{eff}=\oint_{\alpha_{-k}} dx\lambda_{eff},
~~~~~~\int_{C_{k}}dx\lambda_{eff}=\int_{C_{-k}}dx\lambda_{eff}
\ee
For the second equation, it is worth to notice that
$C_{k}-C_{-k}=\sum_{j=-k,j\neq 0}^{k} \beta_j$ and
$\oint_{\beta_{j}} dx\lambda_{eff}=-\oint_{\beta_{-j}} dx\lambda_{eff}$.

\EPSFIGURE[ht]{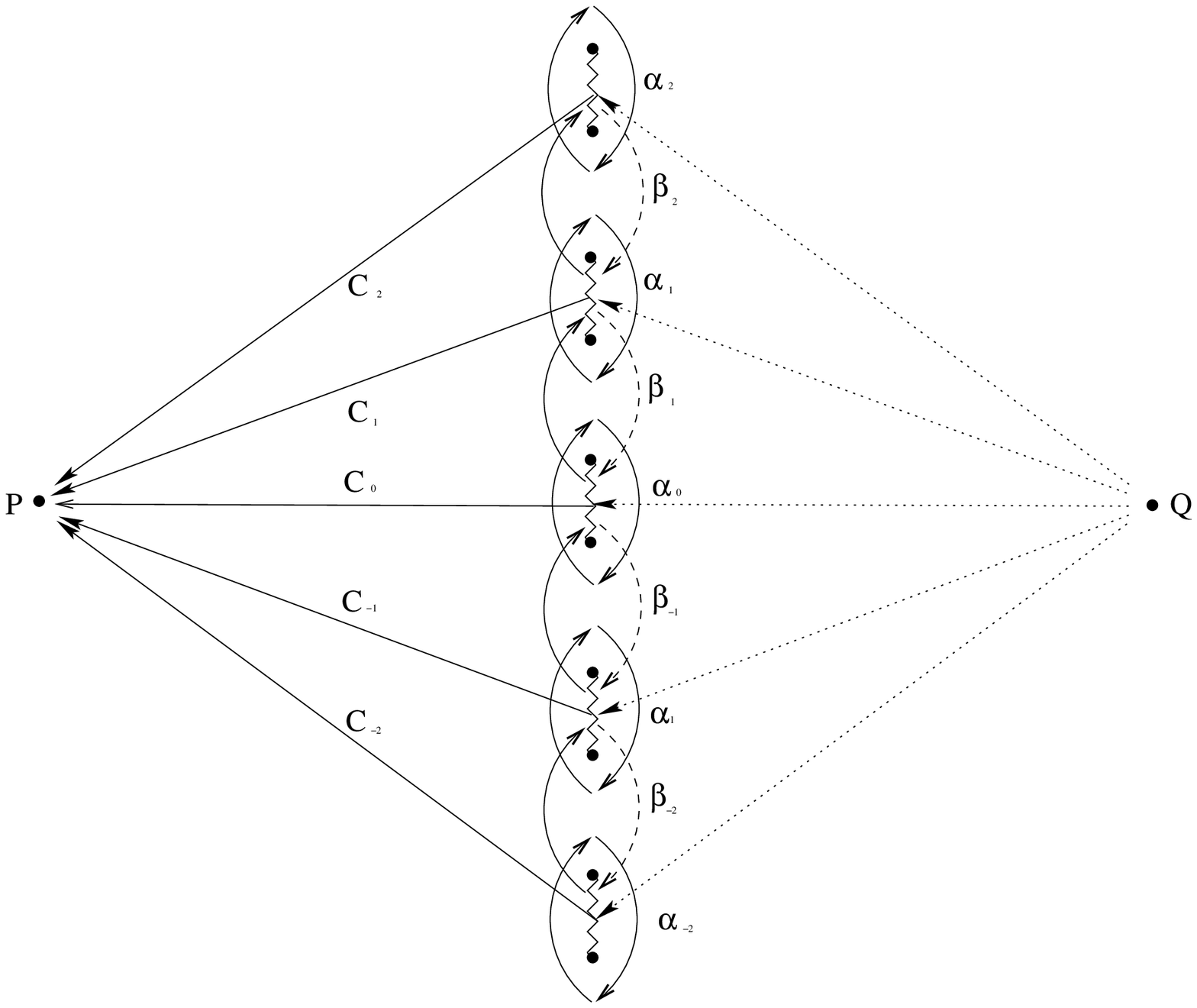,width=11cm}
{The choice of our cycles $\alpha_i,C_i,\beta_i$. Notice that they are 
drawn in symmetric fashion. The solid line means that it is 
at the upper plane while the dotted line, the lower plane. $Q$ is same
point as $P$, but at lower plane.
\label{choice}
}

Now we can identify cycles
$$\int_{a_k}={1\over 2} {1\over 2\pi i}\oint_{\alpha_k},~~~~~~
\int_{b_k}= {1\over 2\pi i}\int_{C_k},$$
and following the physical quantities
\bea
\label{phy_def_S} S_k & = & \int_{a_k}  dx\lambda_{eff}=
{1\over 2} {1\over 2\pi i}\oint_{\alpha_k} dx\lambda_{eff},\\
\Pi_k & =  & {1\over 2}\int_{b_k} dx\lambda_{eff}= 
{1\over 4\pi i}\int_{C_k} dx\lambda_{eff}=
{1\over 2\pi i}\int_{a_k}^{\Lambda_0} dx\lambda_{eff}. \label{phy_def_Pi}
\eea
Among these variables in (\ref{phy_def_S}) and (\ref{phy_def_Pi}),
 (\ref{inte_rel}) indicates that 
\be \label{phy_rel} 
S_{k}=S_{-k},~~~~~~\Pi_{k}=\Pi_{-k}.
\ee
Furthermore, using the fact that D5-branes have been replaced by fluxes,
we get
\be  \label{flux_inte}
\int_{a_k} h={1\over 2} N_i,(k\neq 0),~~~~~~\int_{a_0} h={1\over 2}
2\hat{N}_0,
~~~~~~\int_{b_k} h=2\tau_{YM}~.
\ee
There are several things  to be remarked. First, because of the 
orientifold plane, every D5-brane in the covering space carries
only half of physical brane charges. Secondly, the physical 
brane charges of orientifold
planes are $O5^-=-1$, $O5^+=+1$ and $\widetilde{O5}^-=-1/2$, so 
$2N_0$, which mean initially total $2N_0$  D5-branes wrapped around the
origin, 
are modified to $2N_0-2$ for $SO(2N)$ group, $2N_0-1$ for 
$SO(2N+1)$ group, and $2N_0+2$ for $Sp(2N)$ group.  
Thirdly, 
$\int_{b_k} h$ are  independent of $k$, thus we require
\be \label{h_con_1} 
\int_{C_k-C_l} h=0  \Leftrightarrow \oint_{\beta_j} h=0, \forall j~.
\ee
Fourthly, summing all $\alpha_k$ contours together, we get
\be  \label{h_con_2}
\oint_P {h\over 2\pi i} =\sum_{k=-n}^{n} \oint_{\alpha_k} {h\over 2\pi i} = 
\sum_{k=-n}^{n} N_i=2\hat{N},~~~~\oint_Q {h\over 2\pi i} =-2\hat{N}
\ee
where $2\hat{N}$ is $2N-2$ for $SO(2N)$ gauge group, $2N-1$, 
$SO(2N+1)$ gauge group and $2N+2$, $Sp(2N)$ gauge  group. 
Later, we will find an 1-form $h$ on the surface (\ref{red_Rie})
satisfied both (\ref{h_con_1}) and (\ref{h_con_2}). Finally
putting every thing together we can write the effective action as
\bea
{-1\over 2\pi i} W_{eff}& = &
\sum_{i=-n}^{n} \int_{a_i} h \int_{b_i} dx\lambda_{eff}
 -\int_{b_i} h
\int_{a_i} dx\lambda_{eff}   \nonumber \\
& = & \sum_{i=-n}^n {1\over 2} N_i (2\Pi_i)-2\tau_{YM} S_i \nonumber \\
& = & 2\hat{N}_0 \Pi_0 +(\sum_{i=1}^n 2N_i \Pi_i)-2\tau_{YM} (S_0+2
\sum_{i=1}^n S_i)  \label{reduced_w}
\eea
where in the last line, we keep  cycle integrations of upper
half planes only. 

\subsection{Some properties of $W_{eff}$}
Now let us discuss some properties of the effective action  $W_{eff}$.
First if we let  $\Lambda_0 \rightarrow e^{2\pi i} \Lambda_0$ at the upper
plane, for every $C_k$, it will anti-clockwised enclose all brunch cuts, so 
\bean
\Delta \Pi_k=\Delta ({1\over 2\pi i}\int_{a_k}^{\Lambda_0} dx\lambda_{eff})
=
  -\sum_{i=-n}^{+n} \oint_{\alpha_i} {dx\over 2\pi i}
\lambda_{eff}=-2\sum_{i=-n}^{n} S_i, \\
\eean
which means that $\Pi_k$ must depend on the cutoff $\Lambda_0$ as
\be \label{cut_dep}
\Pi_k= {-2\over 2\pi i} \sum_{i=-n}^{n} S_i \log \Lambda_0+...
\ee
We can also find the $\Lambda_0$ dependence directly by calculating
\bean
\Pi_k & = & {1\over 2\pi}\int_{a_k}^{\Lambda_0} dx\lambda_{eff}  
 =  {1\over 2\pi i}\int_{a_k}^{\Lambda_0} dx
\sqrt{W'(x)^2+f_{2n}(x)} \\
& \sim & {1\over 2\pi i}\int_{a_k}^{\Lambda_0} dx 
(W'(x)+ {b_{2n} \over 2 g_{2n+2}}
{1\over x} +{\cal O}({1\over x^2}) )\\
& \sim & {1\over 2\pi i}( W(\Lambda_0) +{b_{2n} \over 2 g_{2n+2}}  
\log \Lambda_0+{\cal O}({1\over \Lambda_0^2})) \\
\eean
From these two expressions we get a very important relationship
\be \label{S_b}
{-b_{2n} \over 4 g_{2n+2}} = \sum_{i=-n}^{n} S_i
\ee
In fact, this result can be obtained by summing all $\alpha_k$ cycles
on the upper plane and push them to go around point $P$
\bean
 \sum_{i=-n}^{n} S_i& = & 
\sum_{i=-n}^{+n} \oint_{\alpha_i} {1\over 4\pi i}\lambda_{eff}
 =   {1\over 4\pi i} \oint_{P} dx  \sqrt{W'(x)^2+f_{2n}(x)} \\
& = &  {1\over 4\pi i} \oint_{P} dx (W'(x)+{f_{2n}(x) \over 2 W'(x)}+...)\\
& = &  {1\over 4\pi i} \oint_{P} dx {b_{2n} \over 2g_{2n+2} x} 
 =  {-b_{2n} \over 4g_{2n+2}}
\eean
Notice that at the last step, we integrate around the point at infinite.

Now we put (\ref{cut_dep}) back into the effective action and get
\bean
 W_{eff}& = & 2(\sum_{i=-n}^n S_i) [2\hat{N}\log \Lambda_0+2\pi i \tau_{YM}]+... 
\eean
Absorbing the cutoff $\Lambda_0$ into the physical scale $\Lambda$
by $2\hat{N} \log\Lambda_0+2\pi i \tau_{YM}=2\hat{N}\log \Lambda$, we get an
important result 
\be
\label{W_scale_Geo} 
{d W_{eff} \over d \log \Lambda^{4\hat{N}}}
=\sum_{i=-n}^n S_i ={-b_{2n} \over 4 g_{2n+2}}
\ee
This equation is same as (\ref{W_scale_SOe}), (\ref{W_scale_SOo}) and
(\ref{W_scale_Sp}) got by calculations in the field theory if the
$f_{2n}(x)$ in the field theory side is identified to  the one in the dual 
geometry. We will show it is true.

Using the result (\ref{S_b}) we can rewrite  (\ref{reduced_w}) into
\bean
{-1\over 2\pi i} W_{eff}=\sum_{k=-n}^{n} N_k \int_{a_k}^{\Lambda_0}
{dx \over 2\pi i} \lambda_{eff}
+{b_{2n} \tau_{YM}\over 2 g_{2n+2}}
\eean
In the $\Lambda\rightarrow 0$ limit, $f_{2n}(x)=0$ as well as $b_{2n}=0$.
Thus we have
\be \label{W_geo_cal}
W_{eff} =-\sum_{k=-n}^{n} N_k \int_{a_k}^{\Lambda_0} dx W'(x)
=\sum_{k=-n}^{n} N_k W(a_k)-2\hat{N}W(\Lambda_0)
\ee 
It is equal to the  result (\ref{W_field_SO}) in the field theory  up to a
constant\footnote{There is a small difference  between (\ref{W_field_SO})
and (\ref{W_geo_cal}). In (\ref{W_geo_cal}) $N_0$ are modified to 
$\hat{N}_0$. However, in $\Lambda\rightarrow 0$ limit, $a_0=0$ and
$W(a_0)=0$ so
this difference does not effect anything.}.

Just like  \cite{Cachazo:2002pr}, here we have 
shown that for $SO/Sp$ gauge groups, the $W^f_{eff}$ calculated
in the field theory and the $W^G_{eff}$ calculated in the 
dual geometry have the same
value at the classical limit $\Lambda\rightarrow 0$ and follow the same 
differential equation with respect to $\Lambda$, so they must be the same.
These  results finish the first step of proofs 
that the field theory is equivalent
to the dual geometry.

\subsection{The  $h$ and Related Seiberg-Witten curve}
To show the equivalence between the field theory and the dual geometry
we need to find the one form $h$ satisfied conditions 
(\ref{h_con_1}) and (\ref{h_con_2}). To do so, first we rewrite the
effective action as
\bean
{-2\over 2\pi i} W_{eff}& = & \sum_{k=-n}^n \oint_{\alpha_k}
{h\over 2\pi i} \int_{C_k} {dx\lambda_{eff} \over 2\pi i}
-\oint_{\alpha_k}{dx\lambda_{eff} \over 2\pi i} \int_{C_k}{h\over 2\pi i}\\
& = & 2\hat{N} \int_{C_0} {dx\lambda_{eff} \over 2\pi i}
+{b_{2n}\tau_{YM} \over 2 g_{2n+2}} - \sum_{k=1}^n 2N_k (\sum_{j=1}^k 
\oint_{\beta_{j}} {dx\lambda_{eff} \over 2\pi i})~.
\eean
To reach the last step, we have used following facts
$C_{-k}=\sum_{j=1}^k \beta_{-k}+C_0$, 
$C_{k}=-\sum_{j=1}^k \beta_{k}+C_0$ and 
(\ref{h_con_1}), (\ref{h_con_2}). From this we get 
equations of motion 
\bea \label{equation_motion}
{-2\over 2\pi i}{\partial  W_{eff} \over \partial b_{2l}}
& = &  2\hat{N} \int_{C_0}{dx \over 2\pi i}{\partial \lambda_{eff}
\over \partial b_{2l}}+{\tau_{YN} \over 2 g_{2n+2}}\delta_{l,n}
-\sum_{k=1}^n 2N_k (\sum_{j=1}^k 
\oint_{\beta_{j}} {dx \over 2\pi i}{\partial \lambda_{eff}
\over \partial b_{2l}})
\eea
For $l=n$, since ${\partial \lambda_{eff}\over \partial b_{2n}}\rightarrow
{1\over 2 g_{2n+2}} {dx \over x}$ at large $x$ limit, the first two
terms give ${2\hat{N} \log\Lambda_0 +2\pi i \tau_{YM} \over g_{2n+2}}$ 
with a cutoff
$\Lambda_0$ and we need to introduce some $\Lambda$ (depending on
$b_{2r}$) to satisfy the equation.

For $l<n$, notice that
\bean
dx{\partial \lambda_{eff} \over \partial b_{2l}} & = & { x^{2l} dx \over
2\sqrt{ W'(x)^2+f_{2n}(x)}}={ x^{2l} dx \over 2 \sqrt{F_{2(2n+1)}(x)}}\\
& = & {t^{l} dt \over 4 \sqrt{ t \widetilde{F}_{2n+1}(t)}},~~~~~
t=x^2,~~\widetilde{F}_{2n+1}(t)=F_{2(2n+1)}(x)
\eean
The equations of motion (\ref{equation_motion}) can be rewritten as
\bea \label{upto}
 \hat{N} \int_{C_0} {t^{l} dt \over \widetilde{y}}
=\sum_{k=1}^n N_k (\sum_{j=1}^k 
\oint_{\beta_{j}}{t^{l} dt \over \widetilde{y}}),~~~\forall l
\eea
with 
\be \label{rel-SW} 
\Gamma:  ~~~\widetilde{y}^2= t \widetilde{F}_{2n+1}(t)~.
\ee
The curve (\ref{rel-SW}) is, in fact, the related Seiberg-Witten
curve after the $Z_2$ quotient from the covering space \cite{Ahn:1997wh}.
Its genus is $n$ which corresponds to the fact there are $N$ $U(1)$ left
in the field theory. The integrand ${u^{l} du \over \widetilde{y}}$ 
$l=0,...,n-1$ are bases of holomorphic one forms on $\Gamma$ and    
equations (\ref{upto}) mean that the left hand side is zero up to
some periods on $\Gamma$. According to  Abel's theorem, there must
be a meromorphic function on $\Gamma$ with divisor $\hat{N}[P-Q]$
\footnote{It maybe a little confusing that for $SO(2N+1)$ gauge group
we have $\hat{N}=N-1/2$
not integer. The reason is that from the brane picture, there is a stuck
D5-brane on top of the orientifold without the image, 
so the best way to discuss $SO(2N+1)$
is in the covering space.}. Furthermore, $h$ is a holomorphic
one form on $\Gamma$ with certain properties.

Now we have translated  field theory equations into the existence of 
a particular Riemann surface $\Gamma$. We will show that if the
factorization form holds, the particular Riemann surface
exists. Let us start with $SO(2N)$ gauge group. Rewriting 
\bea \label{A1}
(W'(x)^2+f_{2n}(x)) x^2 H_{2N-2n-2}(x)^2= g_{2n+2}^2 (P_{2N}(x,u)^2-4
\gamma^2  x^4)
\eea
with the boundary condition that $P_{2N}(x,u)|_{\gamma
\rightarrow 0}= x^{2N_0} \prod_{i=1}^n (x^2+a_i^2)^{N_i}$
as
\bean
\widetilde{H}_{N-n-1}(t)^2 \widetilde{y}^2=g_{2n+2}^2 (\widetilde{P}_N(t,u)^2
-4\gamma^2 t^2)
\eean
and  defining 
\be
zt= \widetilde{P}_N(t,u)-{1\over g_{2n+2}} \widetilde{y}(t)
\widetilde{H}_{N-n-1}(t), 
\ee
we get the equation satisfied by $z$
\be \label{SO2Nz}
z-{2\widetilde{P}_N(t,u) \over t}+{4\gamma^2 \over z}=0
\ee
(please notice that ${2\widetilde{P}_N(t,u) \over t}=
{2P_{2N}(x,u) \over x^2}$, so it is the polynomial of $t$). 
Notice that  at this moment the $\gamma$ is an undetermined parameter 
which  will be shown to be the  dynamical scale in the
Seiberg-Witten curve by independent derivations.
From (\ref{SO2Nz}), we see immediately that $z$ has zeros of order $N-1$ at
$P$ and poles   of order $N-1$ at $Q$ and holomorphic elsewhere. Thus
\be \label{define_h}
h= {-dz\over z}
\ee
satisfies all conditions required by the geometry. To check this
we lift to the covering space by replacing $t=x^2$. As noticed in
\cite{Cachazo:2002pr},  integration around $\alpha_k$ cycles does not
depend on $\gamma$, so we can evaluate them by setting $\gamma\rightarrow 0$
\bean
\oint_{\alpha_i} {1\over 2\pi i}h & = & {1\over 2\pi i} 
\oint_{\alpha_i}{-dz\over z} 
 =  {1\over 2\pi i} \oint_{\alpha_i} -d(\log z) \\
& = & {1\over 2\pi i} \oint_{\alpha_i} d({2P_{2N}(x,u) \over x^2}|_{\gamma
\rightarrow 0}) =  N_i {1\over 2\pi i} \oint_{\alpha_i}
-d(\log(x-i a_i)) \\ & = & N_i~~for~(i\neq 0),~~~~~or~(2N_0-2),~i=0
\eean
where we have used the boundary condition  $P_{2N}(x,u)|_{\gamma
\rightarrow 0}= x^{2N_0} \prod_{i=1}^n (x^2+a_i^2)^{N_i}$
and the direction of cycles is clockwise. Furthermore
\bean
\int_{C_i-C_j}{1\over 2\pi i} {-dz\over z} =0
\eean
since $C_i-C_j$ cycles do not cross any branch cut of the logarithmic
function. To determine the $\gamma$, we solve
\be \label{solve_z}
z={P_{2N}(x,u) \over x^2}\pm \sqrt{({P_{2N}(x,u) \over x^2})^2-4 \gamma^2}
\ee
and integrate directly
\bean
2\tau_{YM}=\int_{C_k} {1\over 2\pi i}h 
& = & {2\over 2\pi i}\int_{a_k^{+}}^{\Lambda_0}
{-dz\over z}= {-2\over 2\pi i} \log(z)|_{a_k^{+}}^{\Lambda_0}
 =  {-2\over 2\pi i} \log{2\Lambda_0^{2N-2} \over \pm 2\gamma}
\eean
where we have used the fact that at $x=a_k^{+}$, $W'(x)^2+f_{2n}(x)=0$,
so by the factorization form we have
$$ ({P_{2N}(x,u) \over x^2})^2=4\gamma^2~.$$
Because  we have required $2\pi i\tau_{YM}+(2N-2)\log\Lambda_0=(2N-2)\log \Lambda$,
it gives immediately
\be \label{A2}
\pm \gamma= \Lambda^{2N-2}~.
\ee
Results (\ref{A1}) and   (\ref{A2}) prove that the complex deformation $f_{2n}(x)$ in the dual 
geometry is same as the $f_{2n}(x)$ in the field theory by 
factorization.

Similar calculations can be done for $SO(2N+1)$ and $Sp(2N)$ gauge groups. For 
$SO(2N+1)$, we take the factorized form
\be
(W'(x)^2+f_{2n}(x)) x^2 H_{2N-2n-2}(x)^2= g_{2n+2}^2 (P_{2N}(x,u)^2-4
\gamma^2  x^2)
\ee
with the boundary condition $P_{2N}(x,u)|_{\gamma
\rightarrow 0} =x^{2N_0} \prod_{i=1}^n (x^2+a_i^2)^{N_i}$
and define $z$ by
\be z- {2P_{2N}(x,u) \over x}+ {4 \gamma^2 \over z}=0 \ee~.
Notice that ${2P_{2N}(x,u) \over x}$ does not have poles at $x=0$.
It is easy to see that $z$ has zeros of order $2N-1$ at $P$ and 
poles of order $2N-1$ at $Q$ (notice that now it is in the covering
space). Defining $h$ as in (\ref{define_h})
and doing same calculations, it is easy to show that $h$ satisfies
all required conditions. Directly integrating $h$ along any $C_k$, 
it can be seen that
\be  \pm \gamma=\Lambda^{2N-1} \ee~.
For $Sp(2N)$ gauge group, we use
\be
(W'(x)^2+f_{2n}(x)) x^2 H_{2N-2n}(x)^2  = 
g_{2n+2}^2 [x^2P_{2N}(x,u)+ 2 \gamma ]^2-4 \gamma^2
\ee
with the boundary condition $P_{2N}(x,u)|_{\gamma
\rightarrow 0}= x^{2N_0} \prod_{i=1}^n (x^2+a_i^2)^{N_i}$ and define 
$z$ by the equation
\be
z+{4\gamma^2 \over z}-2(x^2P_{2N}(x,u)+ 2 \gamma)=0
\ee
with zeros of order $2N+2$ at $P$ and 
poles of order $2N+2$ at $Q$. Using $h$ as in (\ref{define_h}) it is
easy to check all required conditions for $h$ and determine 
\be  \pm \gamma=\Lambda^{2N+2} \ee

\subsection{The coupling constant matrix $\tau_{ij}$}
Now the last piece we need to do is to check that the 
coupling constant matrix 
$$\tau_{ij}={\partial^2 {\cal F} \over \partial S_i \partial S_j}
$$
is indeed given by periods of the reduced Seiberg-Witten curve.
Since we have shown that $S_k=S_{-k}, \Pi_{k}=\Pi_{-k}$, 
there are only $n+1$ independent $S_k$ and $\Pi_k$ which, for
simplicity, can be chosen to be $S_k,\Pi_k$ with $k=0,1,...,n$
with relations
\be
\label{PiS} \Pi_k={\partial {\cal F} \over \partial S_k},~~~k>0,~~~~
\Pi_0=2{\partial {\cal F} \over \partial S_0}~.
\ee
The reason for the second equation is that under the $Z_2$ action, 
$S_0$ is mapped to itself, so the physical glueball field $S_0^f=
S_0/2$ and $\Pi_0={\partial {\cal F} \over \partial S_0^f}$.

We define  new bases of cycles as
\be \label{newcycle}
{\partial \over \partial \widetilde{S}_0} =  {1\over 4\hat{N}}[
4\hat{N}_0 {\partial  \over \partial S_0}+ \sum_{j=1}^n
2N_j {\partial  \over \partial S_j}],~~~
{\partial \over \partial \widetilde{S}_{i>1}} = 
{\partial \over \partial S_{ i}}- {\partial \over \partial S_{i-1}},
~~~~{\partial \over \partial \widetilde{S}_{1}} = 
{\partial \over \partial S_{1}}- 2{\partial \over \partial S_{0}}.
\ee
Then using  equations of motion for the effective action (\ref{reduced_w})
\be
{\partial \over \partial S_k}[2\hat{N}_0 \Pi_0 +(\sum_{i=1}^n 2N_i \Pi_i)-2\tau_{YM} (S_0+2
\sum_{i=1}^n S_i)] =0
\ee
we see immediately 
\bean
\widetilde{\tau}_{00} 
=  {\partial^2 {\cal F} \over \partial {\widetilde{S}_0^2}}
= {2\tau_{YM}\over 4\hat{N}},~~~
\widetilde{\tau}_{0,i\neq 0} 
 =  {\partial^2 {\cal F} \over \partial \widetilde{S}_{  i}
\partial {\widetilde{S}_0}}=0~.
\eean
In fact, $\widetilde{\tau}_{00}$ is the coupling constant of 
central $U(1)$ in double covering
$U(2N)$ gauge group. When we project the $U(2N)$ to $SO/Sp$ gauge groups by
orientifold, the $U(1)$  is broken to global $Z_2$ symmetry
as discussed in \cite{Ahn:1997wh}. For other coupling constants
\bean
\widetilde{\tau}_{ij}  =  {\partial^2 \over \partial \widetilde{S}_i 
\partial \widetilde{S}_j} {\cal F} = {\partial \over \partial \widetilde{S}_i}
(\Pi_j-\Pi_{j- 1})
= {\partial \over \partial \widetilde{S}_i} \int_{C_j-C_{j- 1}} 
\lambda_{eff},~~~i,j\geq 1
\eean
by taking $b_{2r}$ as new independent variables, we have
\bea
\widetilde{\tau}_{ij} & = & \sum_{r=0}^{n-1} {\partial b_{2r} \over
\partial \widetilde{S}_i}
{\partial \over \partial b_{2r}} (\int_{C_j-C_{j- 1}} 
\lambda_{eff})+{\partial b_{2n} \over \partial \widetilde{S}_i}
{\partial \over \partial b_{2n}} (\int_{C_j-C_{j- 1}} 
\lambda_{eff}) \nonumber \\
& = & \sum_{r=0}^{n-1} {\partial b_{2r} \over \partial \widetilde{S}_i}
{\partial \over \partial b_{2r}} (\int_{C_j-C_{j-1}} 
\lambda_{eff}) \label{tauij}
\eea
where the second term drops out because 
$b_{2n}=-4g_{2n+2}(S_0+2\sum_{j=1}^n 2 S_j)$ 
and ${\partial b_{2n} \over \partial \widetilde{S}_i}=0$. Using 
$\lambda_{eff}=\sqrt{W'(x)^2+f_{2n}(x)}$, it is easy to see that 
$$
dx{\partial \lambda_{eff}\over \partial b_{2r}} =dx{x^{2r} \over
2\lambda_{eff}}=dt {t^{r}\over 2 \widetilde{y}(t)} ,~~~r=0,..,n-1$$
where $\widetilde{y}$ is given in (\ref{rel-SW}) to be exactly the
reduced Seiberg-Witten curve. Since $dt {t^{r}\over 2 \widetilde{y}(t)}$
with $r=0,1,...,n-1$ form a bases of holomorphic one forms on 
the reduced Riemann surface $\Gamma$ and $\{\alpha_i,C_i-C_{i-1} \}$
form a basis for $H_1(\Gamma,Z)$, by (\ref{tauij}) $\widetilde{\tau}_{ij}$
are indeed given by the period matrix of $\Gamma$. 
This completes the proof that the effective action and the coupling
constants in the field theory can be equivalently calculated by the
dual geometry using the large $N$ duality. 

Before closing this section, let us remark the role of $z$ defined above.
It can be shown that $x{dz \over z}$ 
is exactly the Seiberg-Witten differential. 
For example, in the case of  $SO(2N)$ gauge group, using (\ref{solve_z})
and $y^2=P_{2N}(x^2,u)^2-4\Lambda^{4N-4} x^4$ we get
\bean
x {dz \over z} & = & x dx ({ ({P_{2N}(x,u) \over x^2})' \over
\sqrt{({P_{2N}(x,u) \over x^2})^2-4 \Lambda^{4N-4}}}) 
= {xdx \over y} [ x^2 ({P'_{2N}(x,u) \over x^2}-2{P_{2N}(x,u) \over x^3})]\\
& = & {xdx \over y} [P_{2N}(x,u)- {1\over 2}P_{2N}(x,u)
{(4 \Lambda^{4N-4} x^4)' \over 4 \Lambda^{4N-4} x^4}]   
\eean
which is indeed the Seiberg-Witten differential \cite{D'Hoker:1996mu}. 
In fact, ${dz \over z}$ is nothing else, but
the eigenvalue distribution function in the corresponding gauge field
theory as emphasized in  \cite{Dijkgraaf:2002pp,Gopakumar:2002wx}. 
Furthermore, in the classical limit $\Lambda \rightarrow 0$, we have
\be \label{classical-limit}
{dz \over z}=dx(1-{2\over x})~.
\ee
The term ${2\over x}$ counts the contribution of the orientifold plane. It is
rather strange that {\sl even in the classical limit the theory knows the 
presence  of the orientifold plane}.

\section{The matrix model}
Recalling the proof of matrix model conjecture for $U(N)$ gauge theory 
with superpotential $W(\Phi)$ given in \cite{Dijkgraaf:2002fc}, the first
step is to show  that from the corresponding matrix model,  the
spectral curve which is same as that in the dual geometry, can be derived.
The second step is to match various integrations along compact and non-compact
cycles at both sides (matrix model side and dual geometric side). The
last step is to show that the relationship among these integrations
are same at both sides. We will follow the same logic here for 
$SO/Sp$ gauge groups. 

The matrix models for the $SO/Sp$ gauge groups have been proposed 
in \cite{Ita:2002kx,Sujay,Janik:2002nz}\footnote{In previous version,
we follow the orthogonal and symplectic ensembles matrix model 
in \cite{Fuji:2002wd}. However, from the point of view of field
theory, it is more natural to use the matrix model proposed
in \cite{Ita:2002kx,Sujay,Janik:2002nz}. Our treatment in
this section will follow these three papers.} 
The  partition function of the 
matrix model is given by
\be
Z={1\over Vol(G)} \int d\Phi exp(-{1\over g_s} \Tr W(\Phi))
\ee
where $\Phi$ is in the adjoint representation of relative groups.
The group measure has been given explicitly in \cite{Ooguri:2002gx,Sujay} for
general matrices. 
For these models,  Feynman diagrams are unoriented double line diagrams
which reflect the nature of $SO/Sp$ gauge groups. Going to the eigenvalue integration
we get
\be
Z\sim \int \prod_i d\lambda_i [\prod_{i<j} (\lambda_i^2-\lambda_j^2)^2]
[\prod_i \lambda_i^2]^{s}  \exp(-{2\over g_s} \sum_{i=1}^M \Tr W(\lambda_i))
\ee 
where $s=0$ for $SO(2M)$ and $s=1$ for $SO(2M+1)/Sp(M)$. Putting the
Vandermonde determinant to the exponential we get
\be \label{S-lambda}
S(\lambda)= -{2\over g_s} \sum_{i=1}^M \Tr W(\lambda_i)-\sum_{i<j}
\log(\lambda_i^2-\lambda_j^2)^2-s\sum_{i=1}^M \log \lambda_i^2~.
\ee 
Saddle point approximation of (\ref{S-lambda}) gives us 
equations of motion of eigenvalues
\be \label{eq-motion}
{1\over g_s} W'(\lambda_i)-2\lambda_i \sum_{j\neq i}{1\over \lambda_i^2-
\lambda_j^2}-{s\over \lambda_i}=0 
\ee 
Define the resolvent to be
\be \label{resolvent}
\omega(x)={-1\over M} \Tr{1\over x-\Phi}={1\over M} \sum_{i=1}^M {2x\over x^2-
\lambda_i^2}
\ee
where we have used the fact that both $\pm \lambda_i$ are eigenvalues
of $\Phi$ for $SO/Sp$ gauge groups. With some algebraic operations
we get
\be  \label{loop-equation}
\omega(x)^2-{1\over M}[ \omega^\prime (x)-{1-2s \over x} \omega(x)]
-{4\over \mu^2} f_{2n}(x) +{2\over \mu} \omega(x) W^\prime (x)=0
\ee
where 
\be  \label{f2n}
f_{2n}(x)= g_s \sum_{i=1}^M {\lambda_i W^\prime (\lambda_i)- x W^\prime (x)
\over x^2 -\lambda_i^2}
\ee
and $\mu=g_s M$ which will be kept  to be constant in the large $M$ limit.
Notice that since $W^\prime (x)$ is an odd function of $x$, $f_{2n}(x)$ will
be an even polynomial of $x$ with degree $2n$. Also the difference 
between $SO(2M)$ and $SO(2M+1)/Sp(M)$ in (\ref{loop-equation})
is counted by the $(1-2s)$ factor 
of ${\cal O}(M^{-1})$ order.

After taking the large $M$ limit, differential equation
 (\ref{loop-equation}) becomes algebraic equation
\be \label{large-M}
\omega(x)^2-{4\over \mu^2} f_{2n}(x) +{2\over \mu} \omega(x) W^\prime (x)=0
\ee
from which, if we define 
\be \label{define-y}
y(x)= {\mu\over 2}( \omega(x) + {W^\prime (x) \over 2})
\ee
 we get the spectral curve
\be  \label{A3}
y^2= W^\prime (x)^2+f_{2n}(x)
\ee
Curve (\ref{A3}) is exactly same form (\ref{red_Rie}) as in previous section. 
$y(x)$ is related to
the force of moving eigenvalues away from their equilibrium positions as
\be
y(\lambda)={g_s\over 2}[ {\partial S(\lambda) \over \partial \lambda}
+{s\over \lambda}]\Longrightarrow {g_s\over 2}{\partial S(\lambda) \over \partial \lambda}|_{large~M}
\ee

Defining the eigenvalue distribution function as 
\be
\rho(\lambda)  =  {1\over M} \sum_{i} \delta(\lambda-\lambda_i),~~~
\int d\lambda \rho(\lambda)=1
\ee
we have
\be
\rho(\lambda) ={1\over 2\pi i}(\omega(\lambda+i0)-\omega(\lambda-i0))
\ee
At large $M$ limit, eigenvalues are clustered around  different critical 
points given by the superpotential $W(\Phi)$ and  filling factors can be
calculated as
\be \label{A4}
{M_k \over M}=\oint_{\alpha_k} d\lambda \rho(\lambda)
\ee
Using the definition of $y$, we can get 
$\rho(\lambda) ={1\over \pi i \mu}(y(\lambda+i0)-y(\lambda-i0))$, so 
\be
M_k={4 M\over \mu}\oint_{\alpha_k}  d\lambda {y(\lambda) 
\over 2\pi i} \Longrightarrow M_k={8\over g_s} S_k
\ee
by comparing with $S_i ={1\over 2}
 \oint_{\alpha_i} d\lambda{1\over 2\pi i}y(\lambda)$. 
Now changing  filling factors by the amount $\Delta M_i$, 
the action is changed to
\bean
\Delta F_{matrix} & = & \Delta M_i \int_{C_i/2} {y(x) \over g_s} = 
{32 \pi i\over g_s^2} \Delta S_i \int_{C_i/2} {y(x) \over 2\pi i}
={32 \pi i\over g_s^2} \Delta S_i \Pi_i
\eean
and we get ${\partial F_{matrix} \over \partial S_i}={32 \pi i\over g_s^2}\Pi_i
$. So to match results in the dual geometry, we just need to identify
\be \label{A5}
F_{matrix}={32 \pi i\over g_s^2} F_{field}
\ee
Equations (\ref{A3}), (\ref{A4}) and  (\ref{A5}) prove the equivalence 
between the matrix model and the dual geometry.

Before ending this section, let us give an important remark.
In \cite{Dijkgraaf:2002dh}
it was suggested that the total contribution to $SO/Sp$ gauge theories
should include both $S^2$ and $RP^2$ diagrams. Using this idea, explicit
calculations have been carried out in \cite{Ita:2002kx} and it was found that
at least up to order ${\cal O}(S^4)$, the whole result can be written as
coming only from $S^2$ diagrams  with modified color number. Later,
a beautiful proof for $SO$ group was given in in \cite{Janik:2002nz}.
These observations are consistent with the result in the dual geometry
where the integration of fluxes $h$ around the origin is modified
by the presence of the orientifold plane.

\section*{Acknowledgements}
This research is supported under the NSF grant {\bf PHY-0070928}. We
owe a lot of thanks to Freddy Cachazo who explained his work carefully
to us and gave a lot of insightful remarks. We also
like to thank  discussions with  Vijay Balasubramanian,
David Berenstein, Joshua Erlich, Yang-Hui He, Min-xin Huang,
Vishnu Jejjala and Asad Naqvi.

\bibliographystyle{JHEP}

\end{document}